\documentclass[11pt,onecolumn]{IEEEtran}

\usepackage{amsmath}
\usepackage{amssymb}
\usepackage[dvips]{graphicx}
\usepackage{setspace}
\renewcommand{\baselinestretch}{1.89}

\newtheorem{lemma:syc}{Lemma}[section]
\newtheorem{lemma:asyc1}[lemma:syc]{Lemma}
\newtheorem{lemma:asyc2}[lemma:syc]{Lemma}

\newtheorem{prop:jitter}{Proposition}[section]

\newtheorem{lemma:IFI}{Lemma}[section]
\newtheorem{lemma:IFI2}[lemma:IFI]{Lemma}
\newtheorem{lemma:syc_MP}[lemma:IFI]{Lemma}
\newtheorem{lemma:asyc1_MP}[lemma:IFI]{Lemma}
\newtheorem{lemma:asyc2_MP}[lemma:IFI]{Lemma}

\begin{document}

\title{Performance Evaluation of Impulse Radio UWB Systems with Pulse-Based Polarity Randomization$^\textrm{\small{1}}$}
\author{Sinan Gezici$^{2,4}$, \textit{Student Member, IEEE}, Hisashi Kobayashi$^{2}$, \textit{Life Fellow, IEEE}, \\H. Vincent Poor$^{2}$, \textit{Fellow, IEEE}, and Andreas F. Molisch$^{3}$, \textit{Senior Member, IEEE}
\\[12pt]To appear in \emph{IEEE Transactions on Signal Processing}}

\footnotetext[1]{This research is supported in part by the
National Science Foundation under grant CCR-99-79361, and in part
by the New Jersey Center for Wireless Telecommunications. Part of
this material was presented at the IEEE Wireless Communications
and Networking Conference 2004.} \footnotetext[2]{Department of
Electrical Engineering, Princeton University, Princeton 08544,
USA, Tel: (609) 258-2798, Fax: (609) 258-2158, email:
\{sgezici,hisashi,poor\}@princeton.edu}
\footnotetext[3]{Mitsubishi Electric Research Labs, 201 Broadway,
Cambridge, MA 02139, USA and also at Department of Electroscience,
Lund University, Box 118, SE-221 00 Lund, Sweden, e-mail:
Andreas.Molisch@ieee.org} \footnotetext[4]{Corresponding author}

\maketitle

\vspace{-2cm}

\begin{abstract}
In this paper, the performance of a binary phase shift keyed
random time-hopping impulse radio system with pulse-based polarity
randomization is analyzed. Transmission over frequency-selective
channels is considered and the effects of inter-frame interference
and multiple access interference on the performance of a generic
Rake receiver are investigated for both synchronous and
asynchronous systems. Closed form (approximate) expressions for
the probability of error that are valid for various Rake combining
schemes are derived. The asynchronous system is modelled as a
chip-synchronous system with uniformly distributed timing jitter
for the transmitted pulses of interfering users. This model allows
the analytical technique developed for the synchronous case to be
extended to the asynchronous case. An approximate closed-form
expression for the probability of bit error, expressed in terms of
the autocorrelation function of the transmitted pulse, is derived
for the asynchronous case. Then, transmission over an additive
white Gaussian noise channel is studied as a special case, and the
effects of multiple-access interference is investigated for both
synchronous and asynchronous systems. The analysis shows that the
chip-synchronous assumption can result in over-estimating the
error probability, and the degree of over-estimation mainly
depends on the autocorrelation function of the ultra-wideband
pulse and the signal-to-interference-plus-noise-ratio of the
system. Simulations studies support the approximate analysis.

\end{abstract}

%\small \textbf{\textit{Keywords---}$\,$Ultra-wideband (UWB),
%impulse radio (IR), multiple access interference (MAI),
%asynchronous systems.} \normalsize

\section{Introduction}

Since the US Federal Communications Commission (FCC) approved the
limited use of ultra-wideband (UWB) technology \cite{FCC},
communications systems that employ UWB signals have drawn
considerable attention. A UWB signal is defined to possess an
absolute bandwidth larger than $500$MHz or a relative bandwidth
larger than 20\% and can coexist with incumbent systems in the
same frequency range due to its large spreading factor and low
power spectral density. UWB technology holds great promise for a
variety of applications such as short-range high-speed data
transmission and precise location estimation.

Commonly, impulse radio (IR) systems, which transmit very short
pulses with a low duty cycle, are employed to implement UWB
systems (\cite{scholtz}-\cite{andy}). In an IR system, a train of
pulses is sent and information is usually conveyed by the position
or the polarity of the pulses, which correspond to Pulse Position
Modulation (PPM) and Binary Phase Shift Keying
(BPSK)$^{5}$\footnotetext[5]{Since IR is a carrierless system, the
only admissible phases are $0$ and $\pi$. Therefore, BPSK becomes
identical to Binary Amplitude-Shift Keying (BASK) in this case.},
respectively. In order to prevent catastrophic collisions among
different users and thus provide robustness against
multiple-access interference, each information symbol is
represented by a sequence of pulses; the positions of the pulses
within that sequence are determined by a pseudo-random
time-hopping (TH) sequence specific to each user \cite{scholtz}.
The number $N_{f}$ of pulses representing one information symbol
can also be interpreted as pulse combining gain.

In ``classical" impulse radio, the polarity of those $N_{f}$
pulses representing an information symbol is always the same,
whether PPM or BPSK is employed (\cite{scholtz}, \cite{gian}).
Recently, pulse-based polarity randomization was proposed, where
each pulse has a random polarity code ($\pm1$) in addition to the
modulation scheme (\cite{eran1}, \cite{sadler}). The use of
polarity codes can provide additional robustness against
multiple-access interference \cite{eran1} and help optimize the
spectral shape according to FCC specifications by eliminating the
spectral lines that are inherent in IR systems without polarity
randomization \cite{paul}.

A TH-IR system with pulse-based polarity randomization can be
considered as a random CDMA (RCDMA) system with a generalized
signature sequence, where the elements of the sequence take values
from $\{-1,0,+1\}$ and are not necessarily independent and
identically distributed (i.i.d.) \cite{eran1}. The performance of
RCDMA systems with i.i.d. binary spreading codes has been
investigated thoroughly in the past (see e.g.
\cite{pursley_1987}-\cite{zang}). Recently,
\cite{ternaryUWB_ciss03} and \cite{ternaryUWB_ciss04} have
considered the problem of designing ternary codes for TH-IR
systems. Moreover, in \cite{eran1}, the performance of random
TH-IR systems with pulse-based polarity randomization has been
investigated over additive white Gaussian noise (AWGN) channels,
assuming symbol-synchronized users. To the best of our knowledge,
no study concerning the bit error probability (BEP) performance of
Rake receivers (with various combining schemes) for a random TH-IR
system with pulse-based polarity randomization in a multiuser,
frequency-selective environment has been reported in the
literature. In this paper, we investigate such a system and
provide (approximate) closed-form expressions for its performance.
We consider an important case in practice, where the different
users are completely asynchronous. We begin by considering the
chip-synchronous case where the symbols of different users are
misaligned but this misalignment is an integer multiple of the
chip interval. Subsequently, we treat a more general asynchronous
case, where we show that the system can be represented as a
chip-synchronous system with uniform timing jitter between zero
and the chip interval for each interfering \emph{user}. We
consider frequency-selective channels and analyze the performance
of Rake receivers with various combining schemes.

\newcounter{sec}
\setcounter{sec}{2} The remainder of the paper is organized as
follows. Section \Roman{sec} describes the transmitted signal
model for a TH-IR system with pulse-based polarity randomization.
In Section \setcounter{sec}{3}\Roman{sec}, both chip-synchronous
and asynchronous systems over frequency-selective channels are
considered, and the performance of Rake receivers is analyzed for
various combining schemes. Simulation studies are presented in
Section \setcounter{sec}{4}\Roman{sec}, followed by some
concluding remarks in Section \setcounter{sec}{5}\Roman{sec}.

\section{Signal Model}

We consider a BPSK random TH-IR system with $N_u$ users, where the
transmitted signal from user $k$ is represented by
\begin{gather}\label{eq:tran1}
s^{(k)}_{tx}(t)=\sqrt{\frac{E_k}{N_f}}\sum_{j=-\infty}^{\infty}d^{(k)}_j\,
b^{(k)}_{\lfloor j/N_f\rfloor}w_{tx}(t-jT_f-c^{(k)}_jT_c),
\end{gather}
where $w_{tx}(t)$ is the transmitted UWB pulse with duration
$T_c$, $E_k$ is the bit energy of user $k$, $T_f$ is the ``frame"
time, $N_f$ is the number of pulses representing one information
symbol, and $b^{(k)}_{\lfloor j/N_f \rfloor}\in \{+1,-1\}$ is the
information symbol transmitted by user $k$. In order to allow the
channel to be shared by many users without causing catastrophic
collisions, a time-hopping sequence $\{c^{(k)}_j\}$ is assigned to
each user, where $c^{(k)}_j \in \{0,1,...,N_c-1\}$ with equal
probability, with $N_c$ denoting the number of possible pulse
positions in a frame ($N_c=T_f/T_c$), and $c^{(k)}_j$ and
$c^{(l)}_i$ are independent for $(k,j)\ne (l,i)$. This TH sequence
provides an additional time shift of $c^{(k)}_jT_c$ seconds to the
$j$th pulse of the $k$th user where the pulse width $T_c$ is also
considered as the chip interval.

$N=N_fN_c$ represents the total processing gain of the system. Due
to the regulations by the FCC \cite{FCC}, each user can transmit a
certain amount of energy in a given time interval. Since the
symbol (bit) energy of the signal defined in (\ref{eq:tran1}) is
constant (denoted by $E_k$), we consider a fixed symbol interval;
hence, a constant total processing gain $N$ throughout the paper.

\begin{figure}
\begin{center}
\includegraphics[width = 0.65\textwidth]{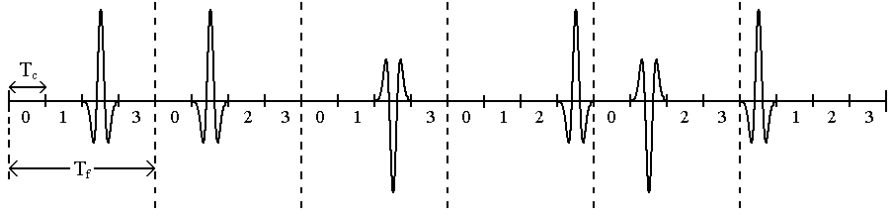}
\caption{A TH-IR signal with pulse-based polarity randomization
where $N_f=6$, $N_c=4$ and the TH sequence is $\{2,1,2,3,1,0\}$.
Assuming that $+1$ is currently being transmitted, the polarity
codes for the pulses are $\{+1,+1,-1,+1,-1,+1\}$.}
\label{fig:codedIR}
\end{center}
\end{figure}

The random polarity codes $d^{(k)}_j$'s are binary random
variables taking $\pm1$ with equal probability, and such that
$d^{(k)}_j$ and $d^{(l)}_i$ are independent for $(k,j)\ne (l,i)$
\cite{eran1}. Use of random polarity codes helps reduce the
spectral lines in the power spectral density of the transmitted
signal \cite{paul} and mitigate the effects of MAI \cite{eran1}.
The receiver for user $k$ is assumed to know its polarity code.

Defining a sequence $\{s^{(k)}_j\}$ as
\begin{gather}\label{eq:spread}
s^{(k)}_j=\begin{cases}d^{(k)}_{\lfloor j/N_c\rfloor} \,\,\,\,
j-N_f\lfloor j/N_c\rfloor=c^{(k)}_{\lfloor j/N_c\rfloor} \\ \quad
0 \quad\quad\quad\quad \textrm{Otherwise}.
\end{cases},
\end{gather}
we can express (\ref{eq:tran1}) as
\begin{equation}\label{tran_CDMA}
s^{(k)}_{tx}(t)=\sqrt{\frac{E_k}{N_f}}\sum_{j=-\infty}^{\infty}s^{(k)}_j\,
b^{(k)}_{\lfloor j/N_fN_c\rfloor}w_{tx}(t-jT_c),
\end{equation}
which indicates that a TH-IR system with polarity randomization
can be regarded as an RCDMA system with a generalized spreading
sequence $\{s^k_j\}$ (\cite{gian2}, \cite{eran1}). Note that the
main difference of the signal model in (\ref{eq:tran1}) from the
``classical" RCDMA model (\cite{pursley_1987}-\cite{zang}) is the
use of $\{-1,0,+1\}$ as the spreading sequence, instead of
$\{-1,+1\}$. The system model given by equation (\ref{eq:tran1})
can represent an RCDMA system with a processing gain of $N_f$, by
considering the special case when $T_f=T_c$.

An example TH-IR signal is shown in Figure \ref{fig:codedIR},
where six pulses are transmitted for each information symbol
($N_f=6$) with the TH sequence $\{2,1,2,3,1,0\}$.

\section{Performance Analysis}

We consider transmission over frequency selective channels, where
the channel for user $k$ is modelled as
\begin{gather}\label{eq:chan_k}
h^{(k)}(t)=\sum_{l=1}^{L}\alpha^{(k)}_l\delta(t-(l-1)T_c-\tau_k),
\end{gather}
where $\alpha_l^{(k)}$ and $\tau_k$ are the fading coefficient of
the $l$th path and the delay of user $k$, respectively; $T_c$ is
the minimum resolvable path interval. We set $\tau_1=0$ without
loss of generality. We assume that the channel characteristics
remain unchanged over a number of symbol intervals, which can be
justified by considering that the symbol duration in a typical
application is on the order of tens or hundreds of nanoseconds,
and the coherence time of an indoor wireless channel is on the
order of tens of milliseconds.

Note that the channel model in (\ref{eq:chan_k}) is quite general
in that it can model any channel of the form
$\sum_{l=1}^{\hat{L}}\hat{\alpha}_l^{(k)}\delta(t-\hat{\tau}_l^{(k)})$
if the channel is bandlimited to $1/T_c$. Thus, each realization
of an arbitrary (and nonuniformly sampled) channel model, e.g.,
the 802.15.3a UWB channel model \cite{Molisch et al. 2003}, can be
represented in the form of equation (4). Only the statistics of
the tap amplitude are changed when the tap spacing is changed to a
uniform spacing.

Using (\ref{eq:tran1}) and (\ref{eq:chan_k}), the received signal
can be expressed as follows:
\begin{align}\label{eq:rec_MP}
r(t)=\sum_{k=1}^{N_u}\sqrt{\frac{E_k}{N_f}}\sum_{j=-\infty}^{\infty}d^{(k)}_j\,
b^{(k)}_{\lfloor
j/N_f\rfloor}u^{(k)}(t-jT_f-c^{(k)}_jT_c-\tau_k)+\sigma_nn(t),
\end{align}
where $n(t)$ is a white Gaussian noise with zero mean and unit
spectral density, and
\begin{gather}\label{eq:u_k}
u^{(k)}(t)=\sum_{l=1}^{L}\alpha_l^{(k)}w_{rx}\left(t-(l-1)T_c\right),
\end{gather}
with $w_{rx}(t)$ being the received UWB pulse with unit energy.

We consider a Rake receiver for the user of interest, say
user $1$, %, which will combine $M$, out of $L$, multipath
%components.
and express the template signal at the Rake receiver as follows:
\begin{gather}\label{eq:temp_RAKE}
s^{(1)}_{temp}(t)=\sum_{j=iN_f}^{(i+1)N_f-1}d_j^{(1)}v(t-jT_f-c_j^{(1)}T_c),
\end{gather}
where
\begin{gather}\label{eq:v}
v(t)=\sum_{l=1}^{L}\beta_lw_{rx}\left(t-(l-1)T_c\right),
\end{gather}
with $\boldsymbol{\beta}=[\beta_1,...,\beta_L]$ being the Rake
combining weights.

The template signal given by (\ref{eq:temp_RAKE}) and (\ref{eq:v})
can represent different multipath diversity combining schemes by
choosing an appropriate weighting vector $\boldsymbol{\beta}$: In
an $M$-finger Rake the weights for $(L-M)$ multipath components
not used in the Rake receiver are set to zero while the remaining
$M$ weights are determined according to the combining scheme, such
as ``Equal Gain Combining (EGC)" or ``Maximum Ratio Combining
(MRC)".

The output of the Rake receiver can be obtained from
(\ref{eq:rec_MP})-(\ref{eq:v}) as follows:
\begin{equation}\label{eq:RAKE_out}
y_1=b_i^{(1)}\sqrt{E_1N_f}\sum_{l=1}^{L}\alpha^{(1)}_l\beta_l+\hat{a}+a+n,
\end{equation}
where the first term is due to the desired signal, $\hat{a}$ is
the self interference of the received signal from user $1$ itself,
which we call inter-frame interference (IFI), $a$ is the MAI from
other users and $n$ is the output noise, which is approximately
distributed as
$n\sim\mathcal{N}\left(0\,,\,N_f\sigma_n^2\sum_{l=1}^{L}\beta_l^2\right)$
for large $N_f$ (Appendix \ref{app:noise}).

\begin{figure}
\begin{center}
\includegraphics[width =0.6\textwidth]{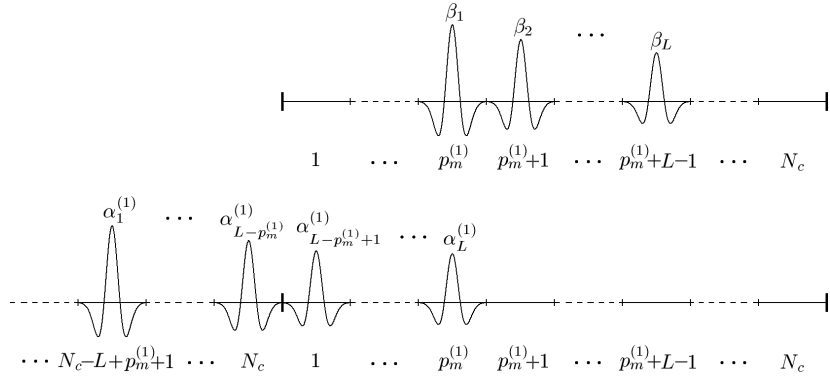}
\caption{Inter-frame interference (IFI) from the $(m-1)$th frame
to the $m$th frame, where $p_m^{(1)}$ denotes the position of the
first user's pulse in the $m$th frame. Only the signals from the
$m$th frame of the template (the signal on the top) and from the
$(m-1)$th frame of the first user are shown. The IFI can also
result from a spill-over of the signal at the $m$th frame of the
template to the $(m+1)$th frame when any of the pulses of the
first user in the $(m+1)$th frame overlap with those pulses that
spill over.} \label{fig:IFI}
\end{center}
\end{figure}

Inter-frame interference (IFI) occurs when a pulse of user $1$ in
a frame spills over to an adjacent frame due to the multipath
effect and consequently interferes with the pulse in that frame
(Figure \ref{fig:IFI}). The IFI in (\ref{eq:RAKE_out}) can be
expressed, from (\ref{eq:rec_MP}) and (\ref{eq:temp_RAKE}), as
\begin{equation}\label{eq:IFI}
\hat{a}=\sqrt{\frac{E_1}{N_f}}\sum_{m=iN_f}^{(i+1)N_f-1}\hat{a}_m,
\end{equation}
where
\begin{gather}\label{eq:IFI_m}
\hat{a}_m=d_m^{(1)}\underset{j\ne
m}{\sum_{j=-\infty}^{\infty}}d^{(1)}_jb^{(1)}_{\lfloor
j/N_f\rfloor}\phi_{uv}^{(1)}\left((j-m)T_f+(c_j^{(1)}-c_m^{(1)})T_c\right),
\end{gather}
with $\phi_{uv}^{(k)}(x)$ denoting the cross-correlation between
$u^{(k)}(t)$ of (\ref{eq:u_k}) and $v(t)$ of (\ref{eq:v}):
\begin{gather}\label{eq:phi}
\phi_{uv}^{(k)}(x)=\int_{-\infty}^{\infty} u^{(k)}(t-x)v(t)dt.
\end{gather}

Note that $\hat{a}_m$ in (\ref{eq:IFI_m}) denotes the IFI due to
the transmitted pulse in the $m$th frame of user $1$, and the sum
of such IFI terms over $N_f$ frames is equal to $\hat{a}$, as seen
in (\ref{eq:IFI}). In Appendix \ref{app:IFI}, we show that these
$N_f$ terms form a 1-dependent sequence$^{6}$\footnotetext[6]{A
sequence $\{X_n\}_{n\in\mathbb{Z}}$ is called a $D$-dependent
sequence, if all finite dimensional marginals
$(X_{n_1},...,X_{n_i})$ and $(X_{m_1},...,X_{m_j})$ are
independent whenever $m_1-n_i>D$.} when $L\leq N_c+1$ and their
sum converges to a Gaussian random variable for a large $N_f$.
This result is summarized in the following lemma:

\begin{lemma:IFI}\label{lem:IFI}
As $N_f\longrightarrow \infty$, the IFI $\hat{a}$ in
(\ref{eq:IFI}) is asymptotically normally distributed as
\begin{equation}\label{eq:lemma_IFI}
\hat{a}\sim\mathcal{N}\left(0\,,\,\frac{E_1}{N_c^2}\sum_{j=1}^{L-1}j
\left[\sum_{l=1}^{L-j}\left(\beta_l\alpha^{(1)}_{l+j}+\alpha^{(1)}_l\beta_{l+j}\right)\right]^2\right),
\end{equation}
for $L\leq N_c+1$.
\end{lemma:IFI}

\textit{Proof:} See Appendix \ref{app:IFI}.

Note that for a Rake receiver with one finger such that
$\beta_1=1$ and $\beta_l=0$ for $l=2,...,L$, the expression
reduces to
$\hat{a}\sim\mathcal{N}\left(0\,,\,\frac{E_1}{N_c^2}\sum_{l=1}^{L-1}l\,(\alpha^{(1)}_{l+1})^2\right)$.

\begin{figure}
\begin{center}
\includegraphics[width = 0.7\textwidth]{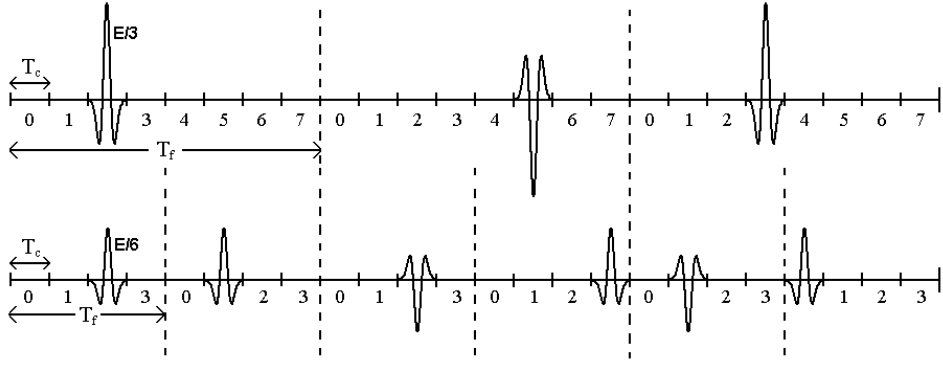}
\caption{Two different cases for a BPSK-modulated TH-IR system
with pulse-based polarity randomization when $N=24$. For the first
case, $N_c=8$, $N_f=3$ and the pulse energy is $E/3$; for the
second case, $N_c=4$, $N_f=6$ and the pulse energy is $E/6$.}
\label{fig:tradeOff}
\end{center}
\end{figure}

Due to the FCC's regulation on peak to average ratio (PAR), $N_f$
cannot be chosen very small in practice. Since we transmit a
certain amount of energy in a constant symbol interval, as $N_f$
gets smaller, the signal becomes more peaky as shown in Figure
\ref{fig:tradeOff}. Therefore, the approximation for large $N_f$
values can be quite accurate for real systems depending on the
system parameters.

When $L>N_c+1$, the pulses in a frame always spill over to the
adjacent frame(s). In this case, the $N_f$  terms in
(\ref{eq:IFI}) form a $\lceil (L-1)/N_c\rceil$-dependent sequence
and the asymptotic distribution of the IFI is given by the
following lemma:

\begin{lemma:IFI2}\label{lem:IFI2}
As $N_f\longrightarrow \infty$, the IFI $\hat{a}$ in
(\ref{eq:IFI}) is asymptotically normally distributed as
\begin{equation}\label{eq:lemma_IFI2}
\hat{a}\sim\mathcal{N}\left(0\,,\,\frac{E_1}{N_c}\sum_{j=1}^{L-N_c}
\left[\sum_{l=1}^{j}\left(\beta_l\alpha^{(1)}_{l+L-j}+\alpha^{(1)}_l\beta_{l+L-j}\right)\right]^2
+\frac{E_1}{N_c^2}\sum_{j=1}^{N_c-1}j
\left[\sum_{l=1}^{L-j}\left(\beta_l\alpha^{(1)}_{l+j}+\alpha^{(1)}_l\beta_{l+j}\right)\right]^2\right),
\end{equation}
for $L>N_c+1$.
\end{lemma:IFI2}

\textit{Proof:} See Appendix \ref{app:IFI2}.

The MAI term in (\ref{eq:RAKE_out}) can be considered as the sum
of MAI terms from each user, that is, $a=\sum_{k=2}^{N_u}a^{(k)}$,
where each $a^{(k)}$ is in turn the sum of interference due to the
signals in the frames of the template:
\begin{equation}\label{eq_MAI_MP}
a^{(k)}=\sqrt{\frac{E_k}{N_f}}\sum_{m=iN_f}^{(i+1)N_f-1}a_m^{(k)},
\end{equation}
with
\begin{gather}\label{eq:MAI_m_MP}
a_m^{(k)}=d_m^{(1)}\sum_{j=-\infty}^{\infty}d^{(k)}_jb^{(k)}_{\lfloor
j/N_f\rfloor}\phi_{uv}^{(k)}\left((j-m)T_f+(c_j^{(k)}-c_m^{(1)})T_c+\tau_k\right),
\end{gather}
where $\phi_{uv}^{(k)}(x)$ is as in (\ref{eq:phi}) and $\tau_k$ is
the delay of the $k$th user.

The effects of MAI will be different for synchronous and
asynchronous systems, as investigated in the following
subsections.

\subsection{Symbol-Synchronous and Chip-Synchronous Cases}

In the symbol-synchronous case, the symbols from different users
are exactly aligned. In other words, $\tau_k=0$ for $k=2,...,N_u$.
On the other hand, for a chip-synchronous scenario, the symbols
are misaligned but the amount of misalignment is an integer
multiple of the chip interval $T_c$. That is,
$\tau_k=\Delta_kT_c$, for $k=2,...,N_u$, where $\Delta_k$ is
uniformly distributed in $\{0,1,...,N-1\}$ with $N=N_cN_f$.

Note that the assumption of synchronism may not be very realistic
for a UWB system due to its high time resolution. However, the aim
of this subsection is two-fold. First, we will show that the BEP
performance of the UWB system is the same whether the users are
symbol-synchronized or chip-synchronized. Second, we will extend
the result for the chip-synchronous case to a more practical
asynchronous case by modelling asynchronous interfering users as
chip-synchronous users with uniform timing jitter, as will be
shown in the next subsection.

The following lemma gives the asymptotic distribution of MAI from
a user for a large number of pulses per symbol.

\begin{lemma:syc_MP}\label{lem:syc_MAI_MP}
As $N_f\longrightarrow\infty$, the MAI from user $k$, which is
chip-synchronized to user $1$, is asymptotically normally
distributed as
\begin{gather}\label{eq:syc_MAI_MP_lemma}
a^{(k)}\sim\mathcal{N}\left(0\,,\,\frac{E_k}{N_c}\left[\sum_{j=1}^{L}\left(\sum_{l=1}^{j}\beta_l\alpha^{(k)}_{l+L-j}\right)^2
+\sum_{j=1}^{L-1}\left(\sum_{l=1}^{j}\alpha^{(k)}_l\beta_{l+L-j}\right)^2\right]\right).
\end{gather}
The result is also valid for a symbol-synchronous scenario.
\end{lemma:syc_MP}

\textit{Proof:} See Appendix \ref{app:syc_MAI_MP}.

Note that when $\beta_1=1$ and $\beta_l=0$, for $l=2,...,L$, we
have
$a^{(k)}\sim\mathcal{N}\left(0\,,\,\frac{E_k}{N_c}\sum_{l=1}^{L}\alpha_l^2\right)$,
which represents the result for a Rake receiver with a single
finger that picks up the first path signal component only.

Note that the Gaussian approximation in Lemma \ref{lem:syc_MAI_MP}
is different from the standard Gaussian approximation (SGA) used
in analyzing a system with many users
(\cite{pursley_1977}-\cite{gardner}). Lemma \ref{lem:syc_MAI_MP}
states that when the number of \textit{pulses} per information
symbol is large, the MAI from an interfering user is approximately
distributed as a Gaussian random variable.

We also note from Lemma \ref{lem:syc_MAI_MP} that the effect of
the MAI is the same for symbol-synchronized and chip-synchronized
cases. This is due mainly to the pulse-based polarity
randomization, which makes the probability distribution of the MAI
independent of the information bits of the interfering user, as
can be observed from (\ref{eq:MAI_m_MP}). Since the probability
that a pulse of the template signal overlaps with any of the
pulses of an interfering user is the same whether the users are
symbol-synchronous or chip-synchronous, the probability
distributions turn out to be the same for both cases.

An approximate expression for BEP can be derived from
(\ref{eq:RAKE_out}), using Lemma \ref{lem:IFI}, Lemma
\ref{lem:IFI2} and Lemma \ref{lem:syc_MAI_MP} as follows:
\begin{equation}\label{eq:BER_syc_MP}
P_{e}\approx
Q\left({\frac{\sqrt{E_1}\sum_{l=1}^{L}\alpha^{(1)}_l\beta_l}{\sqrt{\frac{E_1}{N_cN}\sigma_{IFI,1}^2+\frac{E_1}{N}\sigma_{IFI,2}^2+
\frac{1}{N}\sum_{k=2}^{N_u}E_k\sigma_{MAI,k}^2+\sigma_n^2\sum_{l=1}^{L}\beta_l^2}}}\,\,\right),
\end{equation}
where
\begin{align}\label{eq:sig2_IFI}
\sigma_{IFI,1}^2&=\sum_{j=1}^{\min\{N_c,L\}-1}j
\left[\sum_{l=1}^{L-j}\left(\beta_l\alpha^{(1)}_{l+j}+\alpha^{(1)}_l\beta_{l+j}\right)\right]^2,\\\label{eq:sig2_IFI2}
\sigma_{IFI,2}^2&=\begin{cases}\sum_{j=1}^{L-N_c}
\left[\sum_{l=1}^{j}\left(\beta_l\alpha^{(1)}_{l+L-j}+\alpha^{(1)}_l\beta_{l+L-j}\right)\right]^2,
\quad\quad
L>N_c\\0\,,\quad\quad\quad\quad\quad\quad\quad\quad\quad\quad\quad\quad\quad\quad\quad\quad\quad\quad\quad\,\,\,\,L\leq
N_c
\end{cases},
\end{align}
and
\begin{gather}\label{eq:sig2_MAI_k}
\sigma_{MAI,k}^2=\sum_{j=1}^{L}\left(\sum_{i=1}^{j}\beta_i\alpha^{(k)}_{i+L-j}\right)^2
+\sum_{j=1}^{L-1}\left(\sum_{i=1}^{j}\alpha^{(k)}_i\beta_{i+L-j}\right)^2.
\end{gather}

%From this expression it can be observed that the BEP gets smaller
%for larger processing gain $N$ in an interference-limited
%scenario, since as $N$ increases, the effects of MAI and IFI
%decrease.

Equation (\ref{eq:BER_syc_MP}) implies that, for a fixed total
processing gain $N$, increasing $N_c$, the number of chips per
frame, will decrease the effects of IFI, while the dependency of
the expressions on the MAI remains unchanged. Hence, an RCDMA
system, where $N_f=N$, can suffer from IFI more than any other
TH-IR system with pulse-based polarity randomization, where
$N_f<N$, if the amount of IFI is comparable to the MAI and thermal
noise.

\subsection{Asynchronous Case}

Now consider a completely asynchronous scenario. In this case, it
is assumed that $\tau_k$ in (\ref{eq:MAI_m_MP}) is uniformly
distributed according to $\mathcal{U}[0,NT_c)$ for $k=2,...,N_u$.

In order to calculate the statistics of the MAI term in
(\ref{eq:RAKE_out}), the following simple result will be used.

\begin{prop:jitter}\label{prop1}
The MAI in the asynchronous case has the same distribution as the
MAI in the chip-synchronous case with interfering user $k$ having
a jitter $\epsilon_k$, for $k=2,...,N_u$, which is the same for
all pulses of that user and is drawn from the uniform distribution
$\mathcal{U}[0,T_c)$.
\end{prop:jitter}

\textit{Proof:} Consider (\ref{eq:MAI_m_MP}). For
$k=2,\ldots,N_u$, $\tau_k$ is uniformly distributed in the
discrete set $\{0,T_c,...,(N-1)T_c\}$ in the chip-synchronous
case. In the asynchronous case, $\tau_k$ is a continuous random
variable with distribution $\mathcal{U}[0,NT_c)$. If the jitter
$\epsilon_k$ in the chip-synchronous case is uniformly distributed
with $\mathcal{U}[0,T_c)$, then $\tau_k+\epsilon_k$ is uniformly
distributed as $\mathcal{U}[0,NT_c)$ hence is equivalent to the
distribution of $\tau_k$ in the asynchronous case.$\,\square$

Proposition \ref{prop1} reduces the performance analysis of
asynchronous systems to the calculation of the statistical
properties of
\begin{gather}\label{eq:MAI_m_MP_asy}
a_m^{(k)}=d_m^{(1)}\sum_{j=-\infty}^{\infty}d^{(k)}_jb^{(k)}_{\lfloor
j/N_f\rfloor}\phi_{uv}^{(k)}\left((j-m)T_f+(c_j^{(k)}-c_m^{(1)})T_c+\tau_k+\epsilon_k\right),
\end{gather}
where $\tau_k=\Delta_kT_c$ takes on the values
$\{0,T_c,...,(N-1)T_c\}$ with equal probabilities and
$\epsilon_k\sim\mathcal{U}[0,T_c)$. This problem is similar to the
analysis of TH-IR systems in the presence of timing jitter, which
is studied in \cite{sinan2}. However, in the present case, the
timing jitter of all pulses of an interfering user is the same
instead of being i.i.d.

The following lemma approximates the distribution of the MAI from
an asynchronous user, conditioned on the timing jitter of that
user when the number of pulses per symbol, $N_f$, is large.

\begin{lemma:asyc1_MP}\label{lem:asyc1_MAI_MP}
As $N_f\longrightarrow\infty$, the MAI from user $k$ given
$\epsilon_k$ has the following asymptotic distribution:
\begin{gather}\label{eq:asyc1_MAI_MP_lemma}
a^{(k)}|\epsilon_k\sim\mathcal{N}\left(0\,,\,\frac{E_k}{N_c}\sigma^2_{MAI,k}(\epsilon_k)\right),
\end{gather}
where
\begin{align}\label{eq:MAI_MP_jit}\nonumber
\sigma^2_{MAI,k}(\epsilon_k)&=\sum_{j=0}^{L-1}\left(\sum_{l=1}^{j}\beta_l[\alpha^{(k)}_{l+L-j-1}R(T_c-\epsilon_k)+\alpha^{(k)}_{l+L-j}R(\epsilon_k)]+\beta_{j+1}\alpha^{(k)}_{L}R(T_c-\epsilon_k)\right)^2\\
&+\sum_{j=0}^{L-1}\left(\sum_{l=1}^{j}\alpha^{(k)}_l[\beta_{l+L-j-1}R(\epsilon_k)+\beta_{l+L-j}R(T_c-\epsilon_k)]+\alpha_{j+1}^{(k)}\beta_LR(\epsilon_k)\right)^2,
\end{align}
with $R(x)=\int_{-\infty}^{\infty}w_{rx}(t-x)w_{rx}(t)dt$.
\end{lemma:asyc1_MP}

\textit{Proof:} See Appendix \ref{app:asyc1_MAI_MP}.

Note that when $\epsilon_k=0$, which corresponds to the
chip-synchronized case, (\ref{eq:MAI_MP_jit}) reduces to
(\ref{eq:sig2_MAI_k}).

From Lemma \ref{lem:asyc1_MAI_MP}, we can calculate, for large
$N_f$, an approximate conditional BEP given
$\boldsymbol{\epsilon}=[\epsilon_2\ldots\epsilon_{N_u}]$ as
\begin{equation}\label{eq:BER_MP_jit}
P_{e}|\boldsymbol{\epsilon}\approx
Q\left({\frac{\sqrt{E_1}\sum_{l=1}^{L}\alpha^{(1)}_l\beta_l}{\sqrt{\frac{E_1}{N_cN}\sigma_{IFI,1}^2+\frac{E_1}{N}\sigma_{IFI,2}^2+
\frac{1}{N}\sum_{k=2}^{N_u}E_k\sigma_{MAI,k}^2(\epsilon_k)+\sigma_n^2\sum_{l=1}^{L}\beta_l^2}}}\,\,\right),
\end{equation}
where $\sigma_{MAI,k}^2(\epsilon_k)$ is as in
(\ref{eq:MAI_MP_jit}) and $\sigma_{IFI,1}^2$ and
$\sigma_{IFI,2}^2$ are as in (\ref{eq:sig2_IFI}) and
(\ref{eq:sig2_IFI2}), respectively.

By taking the expectation of (\ref{eq:BER_MP_jit}) with respect to
$\boldsymbol{\epsilon}=[\epsilon_2,\ldots,\epsilon_{N_u}]$, where
$\epsilon_k\sim\mathcal{U}[0,T_c)$ for $k=2,...,N_u$, we find the
BEP:
\begin{gather}\label{eq:accurate_ber}
P_{e}\approx\frac{1}{T_c^{N_u-1}}\int_{0}^{T_c}\ldots\int_{0}^{T_c}
P_{e}|\boldsymbol{\epsilon}\,\,d\epsilon_2\ldots d\epsilon_{N_u}.
\end{gather}

However, when the number of users is large, calculation of
(\ref{eq:accurate_ber}) becomes cumbersome since it requires
integration of $P_{e}|\boldsymbol{\epsilon}$ over $(N_u-1)$
variables. In this case, the SGA
\cite{pursley_1977}-\cite{gardner} can be employed in order to
approximate the BEP in the case of large number of equal energy
interferers:

\begin{lemma:asyc2_MP}\label{lem:asyc2_MAI_MP}
Assume that all the interfering users have the same bit energy
$E$. Then, as $N_u\longrightarrow\infty$, $a/\sqrt{N_u-1}$, where
$a$ is the MAI term in (\ref{eq:RAKE_out}), is asymptotically
normally distributed as
\begin{gather}\label{eq:asyc2_MAI_MP_lemma}
a\sim\mathcal{N}\left(0\,,\,\frac{E}{N_c}\textrm{E}\{\sigma^2_{MAI,k}(\epsilon_k)\}\right),
\end{gather}
where
\begin{align}\label{eq:exp_value}\nonumber
\textrm{E}\{\sigma^2_{MAI,k}(\epsilon_k)\}&=\frac{1}{T_c}\sum_{j=0}^{L-1}\int_{0}^{T_c}\left(\sum_{l=1}^{j}\beta_l[\alpha^{(k)}_{l+L-j-1}R(T_c-\epsilon_k)+\alpha^{(k)}_{l+L-j}R(\epsilon_k)]+\beta_{j+1}\alpha^{(k)}_{L}R(T_c-\epsilon_k)\right)^2d\epsilon_k\\
&+\frac{1}{T_c}\sum_{j=0}^{L-1}\int_{0}^{T_c}\left(\sum_{l=1}^{j}\alpha^{(k)}_l[\beta_{l+L-j-1}R(\epsilon_k)+\beta_{l+L-j}R(T_c-\epsilon_k)]+\alpha_{j+1}^{(k)}\beta_LR(\epsilon_k)\right)^2d\epsilon_k.
\end{align}
\end{lemma:asyc2_MP}

\textit{Proof:} See Appendix \ref{app:asyc2_MAI_MP}.

The BEP can be approximated from Lemma \ref{lem:asyc2_MAI_MP} as
\begin{equation}\label{eq:PER_MP_asy}
P_{e}\approx
Q\left({\frac{\sqrt{E_1}\sum_{l=1}^{L}\alpha^{(1)}_l\beta_l}{\sqrt{\frac{E_1}{N_cN}\sigma_{IFI,1}^2+\frac{E_1}{N}\sigma_{IFI,2}^2+
\frac{E}{N}(N_u-1)\textrm{E}\{\sigma^2_{MAI,k}(\epsilon_k)\}+\sigma_n^2\sum_{l=1}^{L}\beta_l^2}}}\,\,\right),
\end{equation}
for large $N_f$ and $N_u$, and for equal energy interferers.

From (\ref{eq:PER_MP_asy}) we make the same observations as in the
synchronous case. Namely, for a given value of the total
processing gain $N=N_cN_f$, the effect of the MAI on the BEP
remains unchanged while the effect of the IFI increases as the
number of chips per frame, $N_c$, decreases. Hence, the IFI could
be more effective for an RCDMA system, where $N_c=1$.

\subsection{Different Rake Receiver Structures}

In the previous derivations, we have considered a Rake receiver
with $L$ fingers, one at each resolvable multipath component (see
(\ref{eq:temp_RAKE}) and (\ref{eq:v})). A Rake receiver combining
all the paths of the incoming signal is called an
\textit{all-Rake} (\textit{ARake}) receiver. Since a UWB signal
has a very large bandwidth, the number of resolvable multipath
components is usually very large. Hence, an ARake receiver is not
implemented in practice due to its complexity. However, it serves
as a benchmark for the performance of more practical Rake
receivers. A feasible implementation of diversity combining can be
obtained by a \textit{selective-Rake} (\textit{SRake}) receiver,
which combines the $M$ best, out of $L$, multipath components.
Although an SRake receiver is less complex than an ARake receiver,
it needs to keep track of all the multipath components and choose
the best subset of multipath components before feeding it to the
combining stage. A simpler Rake receiver, which combines the first
$M$ paths of the incoming signal, is called a
\textit{partial-Rake} (\textit{PRake}) receiver \cite{cassiICC02}.

The BEP expressions derived in the previous subsections for
synchronous and asynchronous cases are general since one can
express different combining schemes by choosing appropriate
combining weight vector, $\boldsymbol{\beta}$. For example, if we
consider the maximum ratio combining (MRC) scheme, the weights can
be expressed as follows for ARake, SRake and PRake receivers:
%$^5$\footnotetext[5]{Since fading coefficients are
%modelled as real numbers, MRC weights will be chosen to be equal
%to fading coefficients.}.

\subsubsection{ARake}
In this case, the combining weights are chosen as
$\boldsymbol{\beta}=\boldsymbol{\alpha}^{(1)}$, where
$\boldsymbol{\beta}=[\beta_1\ldots\beta_L]$ are the Rake combining
weights in (\ref{eq:v}) and
$\boldsymbol{\alpha}^{(1)}=[\alpha^{(1)}_1\ldots\alpha^{(1)}_L]$
are the fading coefficients of the channel for user $1$.

\subsubsection{SRake}
An SRake receiver combines the best $M$ paths of the received
signal. Let $\mathcal{S}$ be the set of indices of these best
fading coefficients with largest amplitudes. Then, the combining
weights $\boldsymbol{\beta}$ in (\ref{eq:v}) are chosen as
follows:
\begin{gather}
\beta_l=\begin{cases}\alpha_{l}^{(1)},\quad l\in\mathcal{S}\\
0,\quad\quad\,l\notin\mathcal{S}
\end{cases}.
\end{gather}

\subsubsection{PRake}
A PRake receiver combines the first $M$ paths of the received
signal. Therefore, the weights of an SRake receiver with MRC
scheme are given by the following:
\begin{gather}
\beta_l=\begin{cases}\alpha_{l}^{(1)},\quad l=1,\ldots,M\,\\
0,\quad\quad\,\, l=M+1,\ldots,L
\end{cases},
\end{gather}
where $M<L$.

\subsection{Special Case: Transmission over AWGN Channels}

From the analysis of frequency-selective channels, we can obtain
the expressions for AWGN channels as a special case, which might
be useful for intuitive explanations.

Considering the expressions in (\ref{eq:rec_MP})-(\ref{eq:v}), and
setting $\alpha_1=\beta_1=1$ and $\alpha_l=\beta_l=0$ for
$l=2,...,L$, the output of the matched filter (MF) receiver can be
expressed as
\begin{gather}\label{eq:MF_out}
y_1=\sqrt{E_1N_f}\,b_i^{(1)}+a+n,
\end{gather}
where the first term is the signal part of the output, $a$ is the
multiple-access interference (MAI) due to other users and $n$ is
the output noise, distributed as $n\sim
\mathcal{N}(0,\,N_f\sigma_n^2)$. Note that there is no IFI in this
case since a single path channel is assumed.

The MAI is expressed as $a=\sum_{k=2}^{N_u}a^{(k)}$, where the
distribution of $a^{(k)}$ in the symbol-synchronous and
chip-synchronous cases can be obtained from Lemma
\ref{lem:syc_MAI_MP} as
\begin{gather}\label{eq:MAI_k_syc}
a^{(k)}\sim\mathcal{N}\left(0\,,\,\frac{E_k}{N_c}\right).
\end{gather}
Then, the BEP can be obtained as follows:
\begin{gather}\label{eq:BER_syc}
P_{e}\approx
Q\left(\sqrt{\frac{E_1}{\frac{1}{N}\sum_{k=2}^{N_u}E_k+\sigma_n^2}}\,\,\right),
\end{gather}
where $N=N_cN_f$, which is the total processing gain of the
system. Note from (\ref{eq:BER_syc}) that the BEP depends on $N_c$
and $N_f$ only through their product. Hence, the system
performance does not change by changing the number of symbols per
information symbol $N_f$ and the number of chips per frame $N_c$
as long as $N_cN_f$ is held constant. This is different from the
general case of (\ref{eq:BER_syc_MP}), where the IFI is reduced
for larger $N_c$. Therefore, for AWGN channels, the BEP
performance of a TH-IR system with pulse-based polarity
randomization is the same as the special case of an RCDMA system.

Considering \cite{eran1}, the BEP for TH-IR systems
\textit{without} pulse-based polarity randomization is given by
the following expression for the case of a synchronous environment
with a large number of equal energy interferers:
\begin{gather}\label{eq:uncoded}
P_e\approx
Q\left(\sqrt{\frac{E_1}{(N_u-1)\frac{E}{N}\left(1+\frac{N_f-1}{N_c}\right)+\sigma_n^2}}\right),
\end{gather}
where $E$ is the energy of an interferer.

Comparing (\ref{eq:BER_syc}) and (\ref{eq:uncoded}), we observe
that, for $N_f>1$, the MAI affects a TH-IR system without polarity
randomization more than it affects a TH-IR system with pulse-based
polarity randomization and that the gain obtained by polarity
randomization increases as $N_f$ increases (in an
interference-limited scenario). The main reason behind this is
that random polarity codes make each interference term to a pulse
of the template signal (see (\ref{eq:MAI_m_MP})) a random variable
with zero mean since it can be plus or minus interference with
equal probability. On the other hand, without random polarity
codes, the interference terms to the pulses of the template signal
have the same sign, hence add coherently, which increases the
effects of the MAI.

Note that the effects of the MAI reduce if the UWB system without
pulse-based polarity randomization is in an asynchronous
environment. Because, in such a case, the MAI terms from some of
the pulses add up among themselves while the remaining ones add up
among themselves and the polarities of these two groups are
independent from each other. Hence, the average MAI is smaller
than the symbol-synchronous case but it is still larger than or
equal to the MAI for the UWB system \textit{with} pulse-based
polarity randomization, where the sign of each interference term
is independent (see \cite{sinan3} for the trade-off between
processing gains in TH-IR systems with and without polarity
randomization).

For TH-IR systems with polarity randomization, we can approximate,
using Lemma \ref{lem:asyc2_MAI_MP}, the total MAI in the
asynchronous case for a large number of equal energy interferers
as
\begin{gather}\label{eq:MAI_AWGN_asyc}
a \sim
{\mathcal{N}}\left(0\,,\,(N_u-1)\frac{2E}{N_cT_c}\int_{0}^{T_c}R^2(\epsilon)d\epsilon\right).
\end{gather}

Let
$\gamma=\frac{2}{T_c}\int_{0}^{T_c}R^2(\epsilon)d\epsilon=\frac{1}{T_c}\int_{-T_c}^{T_c}R^2(\epsilon)d\epsilon$.
Then, from (\ref{eq:MAI_AWGN_asyc}), $a \sim
{\mathcal{N}}\left(0\,,\,\gamma(N_u-1)E/N_c\right)$. Note from
(\ref{eq:MAI_k_syc}) that for equal energy interfering users, the
MAI in the symbol/chip-synchronous case is distributed as $a \sim
{\mathcal{N}}\left(0\,,\,(N_u-1)E/N_c\right)$. Hence we see that
the difference between the powers of the MAI terms depends on the
autocorrelation function of the UWB pulse. For example, for the
autocorrelation function of (\ref{eq:R1}) below,
$\gamma\approx0.2$ and symbol/chip-synchronization assumption
could possibly result in an over-estimate of the BEP depending on
the signal-to-interference-pulse-noise ratio (SINR) of the system.

From (\ref{eq:MF_out}) and (\ref{eq:MAI_AWGN_asyc}), the BEP of an
asynchronous system can be approximately expressed as follows:
\begin{equation}\label{eq:BER_asycAWGN}
P_{e}\approx
Q\left(\frac{\sqrt{E_1}}{\sqrt{{(N_u-1)\frac{2E}{NT_c}\int_{0}^{T_c}R^2(\epsilon)d\epsilon+\sigma_n^2}}}\right),
\end{equation}
for large values of $N_u$. Similar to the synchronous case, the
performance is independent of the distribution of $N$ between
$N_c$ and $N_f$. Therefore, the TH-IR system performs the same as
an RCDMA system in this case.

\subsection{Average Bit Error Probability}

In order to calculate the average BEP, the previous expressions
for probability of bit error need to be averaged over all fading
coefficients. That is,
$P_{avg}=\textrm{E}\{P_e(\boldsymbol{\alpha}^{(1)},\ldots,\boldsymbol{\alpha}^{(k)})\}$,
which does not lend itself to simple analytical solutions.
However, this average can be evaluated numerically, or by
Monte-Carlo simulations.

\section{Simulation Results}

\begin{figure}
\begin{center}
\includegraphics[width = 0.5\textwidth]{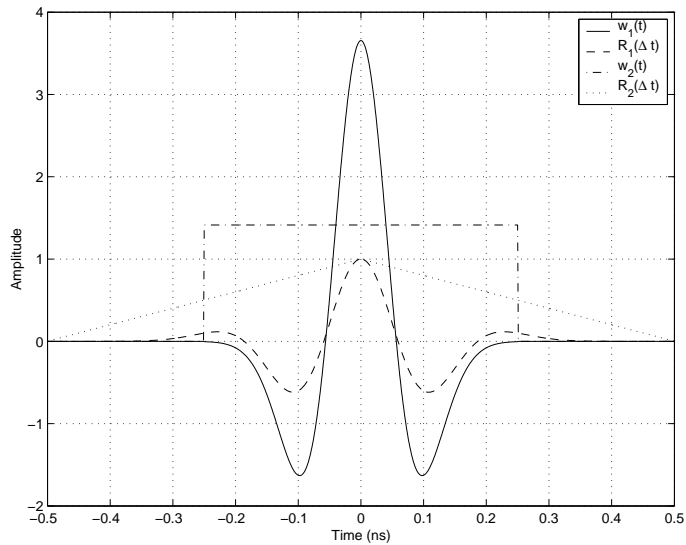}
\caption{UWB pulses and autocorrelation functions for
$T_c=0.5$ns.} \label{fig:pulse_auto}
\end{center}
\end{figure}
In this section, the BEP performance of a TH-IR system with
pulse-based polarity randomization is evaluated by conducting
simulations in MATLAB. The following two types of (unit energy)
UWB pulses and their autocorrelation functions are employed as the
received UWB pulse $w_{rx}(t)$ in the simulations (Figure
\ref{fig:pulse_auto}):
\begin{align}\label{eq:w_1}
w_1(t)&=\left(1-\frac{4\pi t^2}{\tau^2}\right)e^{-2\pi
t^2/\tau^2}/\sqrt{E_p},\\\label{eq:R1} R_1(\Delta
t)&=\left[1-4\pi(\frac{\Delta
t}{\tau})^2+\frac{4\pi^2}{3}(\frac{\Delta
t}{\tau})^4\right]e^{-\pi (\frac{\Delta
t}{\tau})^2},\\\label{eq:w_2} w_2(t)&=\frac{1}{\sqrt{T_c}},\quad
-0.5T_c\leq t \leq 0.5T_c,\\\label{eq:R2} R_2(\Delta
t)&=\begin{cases}-\Delta t/T_c+1,\quad\,\,\, 0\leq \Delta t\leq
T_c\\\Delta t/T_c+1,\quad -T_c\leq \Delta t <0
\end{cases},
\end{align}
where $E_p$ of $w_1(t)$ is the normalization constant,
$\tau=T_c/2.5$ is used in the simulations, and the rectangular
pulse $w_2(t)$ is chosen as an approximate pulse shape in order to
compare the performance of the system with different pulse shapes.

\begin{figure}
\begin{center}
\includegraphics[width = 0.5\textwidth]{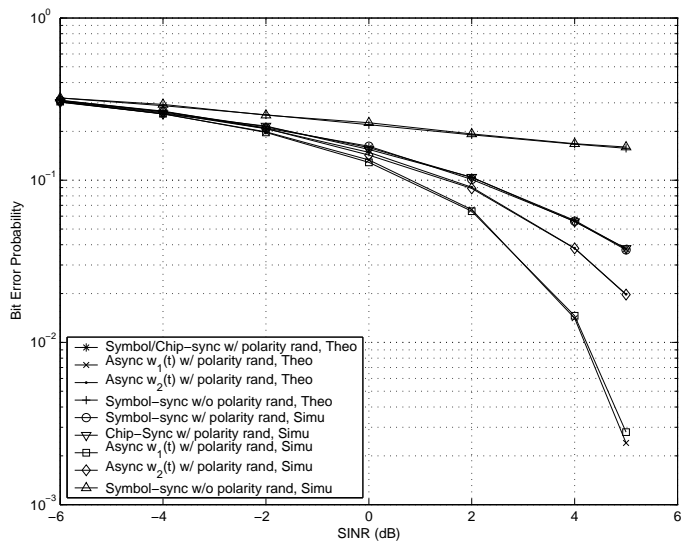}
\caption{BEP vs SINR for different cases, where $N_c=5$, $N_f=15$,
$N_u=10$, $E_1=0.5$ and $E=1$. Transmission over an AWGN channel
is considered.} \label{fig:SIR_vs_BER}
\end{center}
\end{figure}
Figure \ref{fig:SIR_vs_BER} shows the BEP performance of a
$10$-user system ($N_{u}=10$) over an AWGN channel, where
$N_{f}=15$ and $N_{c}=5$. The bit energy of the user of interest,
user $1$, is $E_{1}=0.5$, whereas the interfering users transmit
bits with unit energy ($E_{k}=1$ for $k=2,...,10$), and the
attenuation due to the channel is set equal to unity. The SINR is
defined by
$\textrm{SINR}=10\log_{10}\left({E_1}/{(\frac{1}{N}\sum_{k=2}^{N_u}E_k+\sigma_n^2})\right)$.
In Figure \ref{fig:SIR_vs_BER}, the SINR is varied by changing the
noise power $\sigma_{n}^{2}$ and the BEP is obtained for different
SINR values in the cases of symbol-synchronous, chip-synchronous
and asynchronous TH-IR systems with pulse-based polarity
randomization and a synchronous TH-IR system without pulse-based
polarity randomization$^7$\footnotetext[7]{The results for the
TH-IR system without pulse-based polarity randomization are
provided to justify the discussion in Section III-D. The extensive
comparison between TH-IR systems with and without polarity
randomization is beyond the scope of this paper.}. For the
asynchronous case, performance is simulated for different pulse
shapes $w_{1}(t)$ and $w_{2}(t)$, given by (\ref{eq:w_1}) and
(\ref{eq:w_2}), respectively. From Figure \ref{fig:SIR_vs_BER}, we
see that the simulation results match closely with the theoretical
results. Also note that for small SINR, all the systems perform
quite similarly since the main source of error is the thermal
noise in that case. As the SINR increases, i.e., as the MAI
becomes the limiting factor, the systems start to perform
differently. The asynchronous systems perform better than the
chip-synchronous and symbol-synchronous cases since
$\frac{2}{T_{c}}\int_{0}^{T_{c}}R^{2}(\epsilon)d\epsilon$ in
(\ref{eq:BER_asycAWGN}) is about $0.2$ for $w_{1}(t)$ and $2/3$
for $w_{2}(t)$, which also explains the reason for the lowest bit
error rate of the asynchronous system with UWB pulse $w_{1}(t)$.
Also it is observed that for an IR-UWB system with pulse-based
polarity randomization, the chip-synchronous and the
symbol-synchronous systems perform the same as expected. Moreover,
we observe that without pulse-based polarity randomization, the
MAI is more effective, which results in larger BEP values.

\begin{figure}
\begin{center}
\includegraphics[width = 0.5\textwidth]{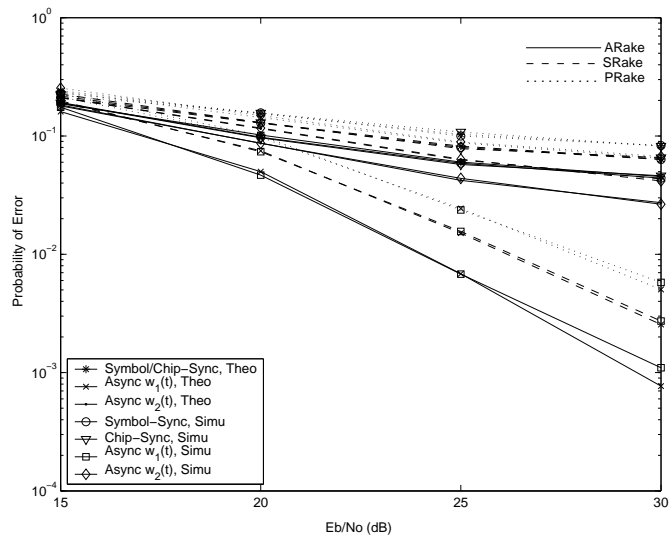}
\caption{Bit error rate vs Eb/No for different cases, where
$N_c=5$, $N_f=15$, $N_u=10$, $E_1=0.5$ and $E=1$. The channel
coefficients are
$[0.4653\,\,0.5817\,\,0.2327-0.4536\,\,0.3490\,\,0.2217-0.1163\,\,0.0233
-0.0116-0.0023]$.} \label{fig:MP1}
\end{center}
\end{figure}

In order to compare the approximate analytical expressions and the
simulation results for multipath channels, we consider the
following channel coefficients for all users:
$\boldsymbol{\alpha}=[0.4653\,\,\,0.5817\,\,\,0.2327\,-0.4536
\,\,\,0.3490\,\,\,0.2217\,-0.1163\,\,\,0.0233\,-0.0116\,-0.0023]$.
Then, the Rake combining fingers are
$\boldsymbol{\beta}=\boldsymbol{\alpha}$ for an ARake receiver,
$\boldsymbol{\beta}=[0.4653\,\,\,0.5817\,\,\,0\,-0.4536
\,\,\,0\,\,\,0\,\,\,0\,\,\,0\,\,\,0\,\,\,0]$ for an SRake receiver
with $3$ fingers, and
$\boldsymbol{\beta}=[0.4653\,\,\,0.5817\,\,\,0.2327\,\,\,0$
$\,\,\,0\,\,\,0\,\,\,0\,\,\,0\,\,\,0\,\,\,0]$ for a PRake receiver
with $3$ fingers. The system parameters are chosen as $N_u=10$,
$N_c=5$, $N_f=15$, $E_1=0.5$ and $E_k=1$ for $k=2,...,10$. Figure
\ref{fig:MP1} plots BEPs of different Rake receivers for
synchronous and asynchronous systems with pulse-based polarity
randomization.
%Note that $E_b/N_0$ is equivalent to $N_f/\sigma_n^2$ in this case since
%each bit is represented by $N_f$ unit energy pulses.
From the figure, we have the same conclusions as in the AWGN
channel case about synchronous and asynchronous cases. Namely,
chip-synchronous and symbol-synchronous systems perform the same
and asynchronous systems with received pulses $w_1(t)$ and
$w_2(t)$ perform better. The asynchronous system with $w_1(t)$
performs the best due to the properties of its correlation
function. Note that the performance is poor when there is
synchronism (chip or symbol level) among the users. However, the
asynchronous system performs reasonably well even in this harsh
multiuser environment. Hence, when computing the BEP of a system,
the assumption of synchronism can result in over-estimating the
BEP. Apart from those, it is also observed from the figure that
the ARake receiver performs the best as expected. Also the SRake
performs better than the PRake since the former collects more
energy because the fourth path is stronger than the third path.

For the next simulations, we model the channel coefficients as
$\alpha_l=\textrm{sign}(\alpha_l)|\alpha_l|$ for $l=1,\ldots,L$,
where $\textrm{sign}(\alpha_l)$ is $\pm1$ with equal probability
and $|\alpha_l|$ is distributed lognormally as
$\mathcal{LN}(\mu_l,\sigma^2)$. Also the energy of the taps is
exponentially decaying as
$\textrm{E}\{|\alpha_l|^2\}=\Omega_0e^{-\lambda(l-1)}$, where
$\lambda$ is the decay factor and
$\sum_{l=1}^{L}\textrm{E}\{|\alpha_l|^2\}=1$ (so
$\Omega_0=(1-e^{-\lambda})/(1-e^{-\lambda L})$).
%This model is in
%compliance with the channel model proposed by IEEE 802.15.3a
%channel modelling subcommittee \cite{IEEE_channel}.
All the system parameters are the same as the previous case,
except we have $E_1=1$ in this case. For the channel parameters,
we have $L=20$, $\lambda=0.25$, $\sigma^2=1$ and $\mu_l$ can be
calculated from
$\mu_l=0.5\left[\textrm{ln}(\frac{1-e^{-\lambda}}{1-e^{-\lambda
L}})-\lambda(l-1)-2\sigma^2\right]$, for $l=1,\ldots,L$.

\begin{figure}
\begin{center}
\includegraphics[width = 0.5\textwidth]{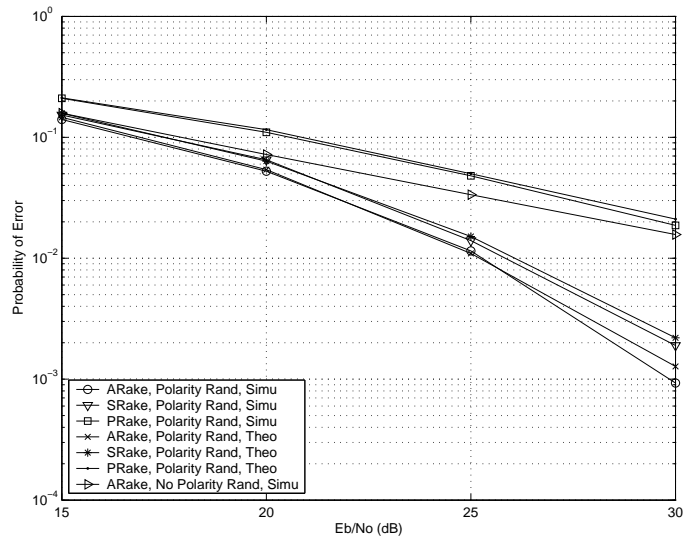}
\caption{Bit error rate vs Eb/No for different receivers in an
asynchronous environment, where $N_c=5$, $N_f=15$, $N_u=10$,
$E_1=1$ and $E=1$. The channel parameters are $L=20$,
$\lambda=0.25$, $\sigma^2=1$. The SRake and PRake have $5$ fingers
each.} \label{fig:MP2}
\end{center}
\end{figure}

Figure \ref{fig:MP2} plots the BEP versus $E_b/N_0$ for different
Rake receivers in an asynchronous environment where $w_1(t)$
models the received UWB pulse. We consider ARake, SRake and PRake
receivers for the TH-IR system with pulse-based polarity
randomization and an ARake receiver for the one without
pulse-based polarity randomization. The SRake and PRake receivers
have $5$ fingers each. As can be seen from the figure, the
theoretical results are quite close to the simulation results.
More accurate results can be obtained when the number of users is
larger. It is also observed that the performance of the SRake
receiver with $5$ fingers is close to that of the ARake receiver
in this setting. Moreover, the ARake receiver for the system
without polarity randomization performs almost as worst as the
PRake receiver for the UWB system with polarity randomization,
which indicates the benefit of polarity randomization in reducing
the effects of MAI.

\begin{figure}
\begin{center}
\includegraphics[width = 0.5\textwidth]{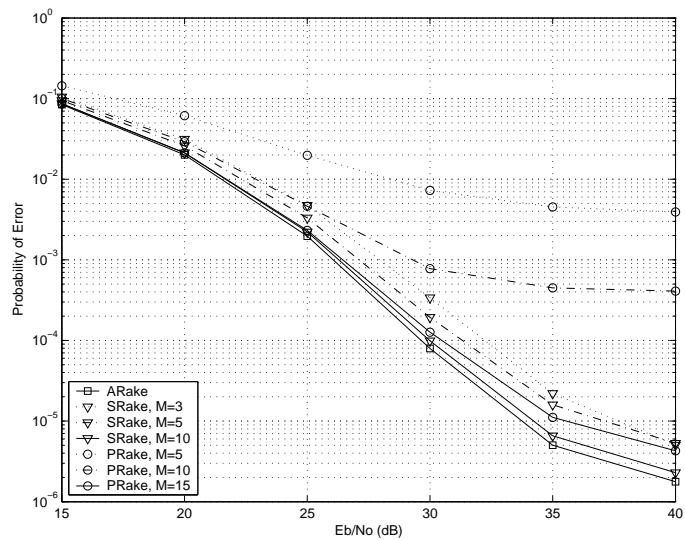}
\caption{Bit error rate vs Eb/No for different receivers in an
asynchronous environment, where $N_c=5$, $N_f=15$, $N_u=10$,
$E_1=2$ and $E=1$. The channel parameters are $L=20$,
$\lambda=0.25$, $\sigma^2=1$.} \label{fig:MP3}
\end{center}
\end{figure}

In Figure \ref{fig:MP3}, we set $E=2$ and keep all the other
parameters the same as in the previous case. Here we consider a
UWB system with polarity randomization and observe the
performances of the SRake and the PRake receivers for different
number of fingers $M$, using (\ref{eq:PER_MP_asy}). It is observed
from the figure that the performance of the SRake receiver with
$10$ fingers is very close to that of the ARake receiver whereas
the PRake receiver needs around $15$ fingers for a similar
performance.

\section{Conclusion}

In this paper, the performance of random TH-IR systems with
pulse-based polarity randomization has been analyzed and
approximate BEP expressions for various combining schemes of Rake
receivers have been derived. Starting from the chip-synchronous
case, we have analyzed the completely asynchronous case by
modelling the latter by an equivalent chip-synchronous system with
uniform timing jitter at interfering users. The effects of MAI and
IFI have been investigated assuming the number of pulses per
symbol is large, and approximate expressions for the BEP have been
derived. Also for a large number of interferers with equal energy,
an approximate BEP expression has been obtained. Simulation
results agree with the theoretical analysis, justifying our
approximate analysis for practical situations.

\appendix
%\section{Appendices}

\subsection{Asymptotic Distribution of $n$ in (\ref{eq:RAKE_out})}\label{app:noise}

The noise term $n$ in (\ref{eq:RAKE_out}) can be obtained from
(\ref{eq:rec_MP}) and (\ref{eq:temp_RAKE}) as $n=\sigma_n\int
s^{(1)}_{temp}(t)n(t)dt$, where $n(t)$ is a zero mean white
Gaussian process with unit spectral density. Hence, $n$ is a
Gaussian random variable for a given template signal. Since the
process has zero mean, $n$ has zero mean for any template signal.
The variance of $n$ can be calculated as
$\textrm{E}\{n^2\}=\sigma_n^2\int(s_{temp}^{(1)}(t))^2dt$ using
the fact that $n(t)$ is white. Using the expressions in
(\ref{eq:temp_RAKE}) and (\ref{eq:v}), we get
\begin{eqnarray}\label{eq:noise_var1}
\textrm{E}\{n^2\}=\sigma_n^2\sum_{j=iN_f}^{(i+1)N_f-1}\int
f_j^2(t)dt+2\sigma_n^2\underset{j\ne
k}{\sum_{j,k=iN_f}^{(i+1)N_f-1}}d^{(1)}_jd^{(1)}_k\int
f_j(t)f_k(t)dt,
\end{eqnarray}
where
$f_j(t)=\sum_{l=1}^{L}\beta_l\,w_{rx}(t-jT_f-c^{(1)}_jT_c-(l-1)T_c-\tau_1)$.

It can be shown that $\int f_j^2(t)dt=\sum_{l=1}^{L}\beta_l^2$ for
all $j$ since $w_{rx}(t)$ is assumed to be a unit energy pulse.
Now consider $\int f_j(t)f_k(t)dt$. By definition, $f_j(t)f_k(t)$
is zero when there is no overlap between the pulses from the $j$th
and the $k$th frames. Assume that $L\leq N_c$. Then,
$f_j(t)f_k(t)=0$ for $|j-k|>1$. In other words, there can be
spill-over from one frame only to a neighboring frame. In this
case, (\ref{eq:noise_var1}) becomes
\begin{eqnarray}\label{eq:noise_var2}
\textrm{E}\{n^2\}=\sigma_n^2N_f\sum_{l=1}^{L}\beta_l^2
+2\sigma_n^2{\sum_{j=iN_f}^{(i+1)N_f-2}}d^{(1)}_jd^{(1)}_{j+1}\int
f_j(t)f_{j+1}(t)dt.
\end{eqnarray}

Note that $f_j(t)$ is a random variable at a given time instant
$t$ due to the presence of the random time-hopping sequence
$\{c_j^{(1)}\}$, and $\{f_j(t)f_{j+1}(t)\}$ are identically
distributed for $j=iN_f,...,(i+1)N_f-2$. Since
$\{d^{(1)}_jd^{(1)}_{j+1}\}$ has zero mean and forms an i.i.d.
sequence for $j=iN_f,...,(i+1)N_f-2$,
$\{d^{(1)}_jd^{(1)}_{j+1}\int f_j(t)f_{j+1}(t)dt\}$ forms a zero
mean i.i.d. sequence. Hence, the second summation in
(\ref{eq:noise_var2}) converge to zero as
$N_f\longrightarrow\infty$, by the Strong Law of Large Numbers.

When the $L\leq N_c$ assumption is removed, we can still use the
same approach to prove the result for finite values of $L$. In
that case, we can write a more general version of
(\ref{eq:noise_var2}) as
\begin{eqnarray}\label{eq:noise_var3}
\textrm{E}\{n^2\}=\sigma_n^2N_f\sum_{l=1}^{L}\beta_l^2
+2\sigma_n^2\sum_{k=1}^{D}{\sum_{j=iN_f}^{(i+1)N_f-1-k}}d^{(1)}_jd^{(1)}_{j+k}\int
f_j(t)f_{j+k}(t)dt,
\end{eqnarray}
where $f_j(t)f_k(t)=0$ for $|j-k|>D$. Since $L$ is assumed to be
finite, $D$ is also finite. Hence, the second term in
(\ref{eq:noise_var3}) still converges to zero as
$N_f\longrightarrow\infty$.

Thus for large $N_f$,
$\textrm{E}\{n^2\}\approx\sigma_n^2N_f\sum_{l=1}^{L}\beta_l^2$,
and so $n$ is approximately distributed as
$n\sim\mathcal{N}\left(0\,,\,\sigma_n^2N_f\sum_{l=1}^{L}\beta_l^2\right)$.

\subsection{Proof of Lemma \ref{lem:IFI}}\label{app:IFI}

The aim is to approximate the distribution of
$\hat{a}=\sqrt{\frac{E_1}{N_f}}\sum_{m=iN_f}^{(i+1)N_f-1}\hat{a}_m$,
where $\hat{a}_m$ is given by (\ref{eq:IFI_m}). Note that
$\hat{a}_m$ denotes the interference to the $m$th frame coming
from the other frames. Assuming that $L\leq N_c+1$, there can be
interference to the $m$th frame only from the $(m-1)$th or
$(m+1)$th frames. Hence, $\hat{a}_m$ can be expressed as:
\begin{gather}\label{eq:IFI_m_ap1}
\hat{a}_m=d_m^{(1)}\sum_{j\in\{m-1,m+1\}}d^{(1)}_jb^{(1)}_{\lfloor
j/N_f\rfloor}\phi_{uv}^{(1)}\left((j-m)T_f+(c_j^{(1)}-c_m^{(1)})T_c\right).
\end{gather}

Note that $\hat{a}_{iN_f},\ldots,\hat{a}_{(i+1)N_f-1}$ are
identically distributed but not independent. However, they form a
1-dependent sequence \cite{meas} since $\hat{a}_{m}$ and
$\hat{a}_{n}$ are independent whenever $|m-n|>1$.

The expected value of $\hat{a}_m$ is equal to zero due to the
random polarity code. That is, $\textrm{E}\{\hat{a}_m\}=0$. The
variance of $\hat{a}_m$ can be calculated from
(\ref{eq:IFI_m_ap1}) as
\begin{gather}\label{eq:var_a_m_ap1}
\textrm{E}\{\hat{a}_m^2\}=\sum_{j\in\{m-1,m+1\}}\textrm{E}\left\{\left[\phi_{uv}^{(1)}\left((j-m)T_f+(c_j^{(1)}-c_m^{(1)})T_c\right)\right]^2\right\},
\end{gather}
where the fact that the random polarity codes are zero mean and
independent for different indices is employed.

Since the TH sequence can take any value in $\{0,1,\ldots,N_c-1\}$
with equal probability, the variance can be calculated as
\begin{gather}\label{eq:var_a_m_ap2}
\textrm{E}\{\hat{a}_m^2\}=\frac{1}{N_c^2}\sum_{j=1}^{L-1}j\left\{[\phi_{uv}^{(1)}(jT_c)]^2+[\phi_{uv}^{(1)}(-jT_c)]^2\right\},
\end{gather}
which can be expressed as
\begin{gather}\label{eq:var_a_m_ap3}
\textrm{E}\{\hat{a}_m^2\}=\frac{1}{N_c^2}\sum_{j=1}^{L-1}j
\left[\left(\sum_{l=1}^{L-j}\beta_l\alpha^{(1)}_{l+j}\right)^2+
\left(\sum_{l=1}^{L-j}\alpha^{(1)}_l\beta_{l+j}\right)^2\right],
\end{gather}
using (\ref{eq:phi}), (\ref{eq:u_k}) and (\ref{eq:v}).

Now consider the correlation terms. Since $L\leq N_c$,
$\textrm{E}\{\hat{a}_m\hat{a}_n\}=0$ when $|m-n|>1$. Hence, we
need to consider $\textrm{E}\{\hat{a}_m\hat{a}_{m+1}\}$ only.
Similar to the derivation of the variance,
$\textrm{E}\{\hat{a}_m\hat{a}_{m+1}\}$ can be obtained, from
(\ref{eq:IFI_m_ap1}), as follows:
\begin{gather}\label{eq:corr_a_m_ap}
\textrm{E}\{\hat{a}_m\hat{a}_{m+1}\}=\frac{1}{N_c^2}\sum_{j=1}^{L-1}j
\left(\sum_{l=1}^{L-j}\beta_l\alpha^{(1)}_{l+j}\right)\left(\sum_{l=1}^{L-j}\beta_l\alpha^{(1)}_{l+j}\right).
\end{gather}

Since $\{\hat{a}_m\}_{m=iN_f}^{(i+1)N_f-1}$ is a zero mean
1-dependent sequence,
$\sqrt{\frac{E_1}{N_f}}\sum_{m=iN_f}^{(i+1)N_f-1}\hat{a}_m$
converges to
\begin{gather}
\mathcal{N}\left(0\,,\,E_1\left[\textrm{E}\{(\hat{a}_m)^2\}+2\textrm{E}\{\hat{a}_m\hat{a}_{m+1}\}\right]\right)
\end{gather}
as $N_f\longrightarrow\infty$ \cite{meas}. Hence,
(\ref{eq:lemma_IFI}) follows from (\ref{eq:var_a_m_ap3}) and
(\ref{eq:corr_a_m_ap}).

\subsection{Proof of Lemma \ref{lem:IFI2}}\label{app:IFI2}

In this section we derive the distribution of IFI for $L>N_c+1$.
Consider the case where $(D-1)N_c+1<L\leq DN_c+1$, with $D$ being
a positive integer. Hence, $\{\hat{a}_m\}_{m=iN_f}^{(i+1)N_f-1}$
forms a $D$-dependent sequence in this case. Similar to Appendix
\ref{app:IFI}, we need to calculate the mean, the variance and the
correlation terms for $\hat{a}_m$ in (\ref{eq:IFI_m}). Due to the
polarity codes, it is clear that $\textrm{E}\{\hat{a}_m\}=0$. The
variance can be expressed as follows, using (\ref{eq:IFI_m}) and
the fact that the polarity codes are zero mean and independent for
different indices:
\begin{gather}\label{eq:appIFI2_var1}
\textrm{E}\{\hat{a}_m^2\}=\underset{j\ne
m}{\sum_{j=-\infty}^{\infty}}\textrm{E}\left\{\left[\phi_{uv}^{(1)}\left((j-m)T_f+(c_j^{(1)}-c_m^{(1)})T_c\right)\right]^2\right\},
\end{gather}
which can be calculated as
\begin{gather}\label{eq:appIFI2_var2}
\textrm{E}\{\hat{a}_m^2\}=\frac{1}{N_c^2}\sum_{i=0}^{N_c-1}\sum_{l=0}^{N_c-1}\underset{j\ne
m}{\sum_{j=-\infty}^{\infty}}\left[\phi_{uv}^{(1)}\left((j-m)T_f+(c_j^{(1)}-c_m^{(1)})T_c\right)\right]^2,
\end{gather}
using that fact that the TH sequence is uniformly distributed in
$\{0,1,\ldots,N_c-1\}$.

Then, the variance term can be expressed as
\begin{gather}\label{eq:appIFI2_var3}
\textrm{E}\{\hat{a}_m^2\}=\frac{1}{N_c^2}\sum_{j=1}^{N_c-1}j\left\{
\left[\phi_{uv}^{(1)}(jT_c)\right]^2+\left[\phi_{uv}^{(1)}(-jT_c)\right]^2\right\}
+\frac{1}{N_c}\sum_{j=N_c}^{L-1}\left\{
\left[\phi_{uv}^{(1)}(jT_c)\right]^2+\left[\phi_{uv}^{(1)}(-jT_c)\right]^2\right\},
\end{gather}
which can be obtained, using (\ref{eq:phi}), (\ref{eq:u_k}) and
(\ref{eq:v}), as follows:
\begin{align}\label{eq:exp2_IFI}\nonumber
\textrm{E}\{(\hat{a}_m)^2\}&=\frac{1}{N_c}\sum_{j=1}^{L-N_c}\left[\left(\sum_{i=1}^{j}\beta_i\alpha^{(1)}_{L+i-j}\right)^2+\left(\sum_{i=1}^{j}\alpha^{(1)}_i\beta_{L+i-j}\right)^2\right]\\
&+\frac{1}{N_c^2}\sum_{j=1}^{N_c-1}j\left[\left(\sum_{i=1}^{L-j}\beta_i\alpha^{(1)}_{i+j}\right)^2+\left(\sum_{i=1}^{L-j}\alpha^{(1)}_i\beta_{i+j}\right)^2\right].
\end{align}

Since $\hat{a}_{iN_f},\ldots,\hat{a}_{(i+1)N_f-1}$ form a
$D$-dependent sequence, we need to calculate
$\textrm{E}\{\hat{a}_m\hat{a}_{m+n}\}$ for $n=1,\ldots,D$. Then,
the IFI term $\hat{a}$ in (\ref{eq:IFI}) can be approximated by
\begin{gather}\label{eq:gaus_IFI2}
{\mathcal{N}}\left(0\,,\,E_1\left[\textrm{E}\{(\hat{a}_{iN_f})^2\}+2\sum_{n=1}^{D}\textrm{E}\{\hat{a}_{iN_f}\hat{a}_{iN_f+n}\}\right]\right),
\end{gather}
as $N_f\longrightarrow\infty$ \cite{meas}.

Using (\ref{eq:IFI_m}), (\ref{eq:phi}), (\ref{eq:u_k}) and
(\ref{eq:v}), the correlation term in (\ref{eq:gaus_IFI2}) can be
calculated, after some manipulation, as

\begin{align}\label{eq:totalCrossIFI}\nonumber
\sum_{n=1}^{D}\textrm{E}\{\hat{a}_{iN_f}\hat{a}_{iN_f+n}\}&=\frac{1}{N_c}\sum_{j=1}^{L-N_c}\left(\sum_{i=1}^{j}\beta_i\alpha^{(1)}_{L+i-j}\right)\left(\sum_{i=1}^{j}\alpha^{(1)}_i\beta_{L+i-j}\right)\\
&+\frac{1}{N_c^2}\sum_{j=1}^{N_c-1}j\left(\sum_{i=1}^{L-j}\beta_i\alpha^{(1)}_{i+j}\right)\left(\sum_{i=1}^{L-j}\alpha^{(1)}_i\beta_{i+j}\right).
\end{align}

Hence, (\ref{eq:lemma_IFI2}) can be obtained by inserting
(\ref{eq:exp2_IFI}) and (\ref{eq:totalCrossIFI}) into
(\ref{eq:gaus_IFI2}).

\subsection{Proof of Lemma \ref{lem:syc_MAI_MP}}\label{app:syc_MAI_MP}

In order to calculate the distribution of the MAI from user $k$,
$a^{(k)}=\sqrt{\frac{E_k}{N_f}}\sum_{m=iN_f}^{(i+1)N_f-1}a_m^{(k)}$,
we first calculate the mean and variance of $a_m^{(k)}$ given by
(\ref{eq:MAI_m_MP}), where the delay of the user, $\tau_k$, is an
integer multiple of the chip interval: $\tau_k=\Delta_kT_c$.

Due to the polarity codes, the mean is equal to zero for any delay
value $\tau_k$; that is, $\textrm{E}\{a_m^{(k)}|\Delta_k\}=0$. In
order to calculate the variance, we make use of the facts that the
polarity codes are independent for different user and frame
indices, and that the TH sequence is uniformly distributed in
$\{0,1,\ldots,N_c-1\}$. Then, we obtain the following expression:
\begin{gather}\label{eq:mai_syc_ap1}
\textrm{E}\{(a_m^{(k)})^2|\Delta_k\}=\frac{1}{N_c^2}\sum_{i=0}^{N_c-1}\sum_{l=0}^{N_c-1}\sum_{j=-\infty}^{\infty}
\left\{\phi_{uv}^{(k)}\left[(i-l+(j-m)N_c+\Delta_k)T_c\right]\right\}^2,
\end{gather}
which is equal to
\begin{gather}\label{eq:mai_syc_ap2}
\textrm{E}\{(a_m^{(k)})^2|\Delta_k\}=\frac{1}{N_c}\sum_{j=-(L-1)}^{L-1}
\left[\phi_{uv}^{(k)}(jT_c)\right]^2.
\end{gather}

Using (\ref{eq:phi}), (\ref{eq:u_k}) and (\ref{eq:v}),
(\ref{eq:mai_syc_ap2}) can be expressed as
\begin{gather}\label{eq:mai_syc_ap3}
\textrm{E}\{(a_m^{(k)})^2|\Delta_k\}=\frac{1}{N_c}\left[\sum_{j=1}^{L}\left(\sum_{i=1}^{j}\beta_i\alpha^{(k)}_{i+L-j}\right)^2
+\sum_{j=1}^{L-1}\left(\sum_{i=1}^{j}\alpha^{(k)}_i\beta_{i+L-j}\right)^2\right].
\end{gather}

Moreover, we note that
$\textrm{E}\{a_m^{(k)}a_n^{(k)}|\Delta_k\}=0$ for $m\ne n$ due to
the polarity codes.

Similar to the proofs in Appendices \ref{app:IFI} and
\ref{app:IFI2}, $\{a_m^{(k)}\}_{m=iN_f}^{(i+1)N_f-1}$ forms a
dependent sequence and the MAI from user $k$,
$a^{(k)}=\sqrt{\frac{E_k}{N_f}}\sum_{m=iN_f}^{(i+1)N_f-1}a_m^{(k)}$,
converge to
${\mathcal{N}}\left(0\,,\,E_k\textrm{E}\{(a_m^{(k)})^2\}\right)$
since the correlation terms are zero. Hence,
(\ref{eq:syc_MAI_MP_lemma}) can be obtained from
(\ref{eq:mai_syc_ap3}).

Note that the result is true for any value of $\Delta_k$ since
$\textrm{E}\{(a_m^{(k)})^2|\Delta_k\}$ in (\ref{eq:mai_syc_ap3})
is independent of $\Delta_k$. Hence, the result is valid for both
symbol and chip synchronous cases.

\subsection{Proof of Lemma \ref{lem:asyc1_MAI_MP}}\label{app:asyc1_MAI_MP}

The proof of Lemma \ref{lem:asyc1_MAI_MP} is an extension of that
of Lemma \ref{lem:syc_MAI_MP}. Considering
(\ref{eq:MAI_m_MP_asy}), we have an additional offset
$\epsilon_k$, which causes a partial overlap between pulses from
the template signal and those from the interfering signal.

Due to the presence of random polarity codes, the mean of
$a_m^{(k)}$ in (\ref{eq:MAI_m_MP_asy}) is equal to zero. Using the
fact that the polarity codes are zero mean and independent for
different frame indices and that the TH codes are uniformly
distributed in $\{0,1,\ldots,N_c-1\}$, we can calculate the
variance of $a_m^{(k)}$ conditioned on $\Delta_k$ and $\epsilon_k$
as
\begin{gather}\label{eq:mai_asyc_ap1}
\textrm{E}\{(a_m^{(k)})^2|\Delta_k,\epsilon_k\}=\frac{1}{N_c^2}\sum_{i=0}^{N_c-1}\sum_{l=0}^{N_c-1}\sum_{j=-\infty}^{\infty}
\left\{\phi_{uv}^{(k)}\left[(i-l+(j-m)N_c+\Delta_k)T_c+\epsilon_k\right]\right\}^2,
\end{gather}
which can be shown to be equal to
\begin{gather}\label{eq:mai_asyc_ap2}
\textrm{E}\{(a_m^{(k)})^2|\Delta_k,\epsilon_k\}=\frac{1}{N_c}\sum_{j=-L}^{L-1}
\left[\phi_{uv}^{(k)}(jT_c+\epsilon_k)\right]^2.
\end{gather}

Note that since the expression in (\ref{eq:mai_asyc_ap2}) is
independent of $\Delta_k$,
$\textrm{E}\{(a_m^{(k)})^2|\Delta_k,\epsilon_k\}=\textrm{E}\{(a_m^{(k)})^2|\epsilon_k\}$.

From (\ref{eq:phi}), (\ref{eq:u_k}) and (\ref{eq:v}), we can
obtain an expression for $\phi_{uv}^{(k)}(jT_c+\epsilon_k)$ when
$j\geq0$ as
\begin{gather}\label{eq:phi1}
\phi_{uv}(jT_c+\epsilon_k)=\sum_{l=1}^{L-j-1}\alpha_l^{(k)}\left[\beta_{l+j}R(\epsilon_k)+\beta_{l+j+1}R(T_c-\epsilon_k)\right]+\alpha_{L-j}^{(k)}\beta_LR(\epsilon_k),
\end{gather}
where $R(x)=\int_{-\infty}^{\infty}w_{rx}(t-x)w_{rx}(t)dt$.
Similarly, the expression for $\phi_{uv}^{(k)}(-jT_c+\epsilon_k)$
can be expressed as follows for $j>0$:
\begin{gather}\label{eq:phi2}
\phi_{uv}(-jT_c+\epsilon_k)=\sum_{l=1}^{L-j}\beta_l\left[\alpha^{(k)}_{l+j}R(\epsilon_k)+\alpha^{(k)}_{l+j-1}R(T_c-\epsilon_k)\right]+\beta_{L-j+1}\alpha^{(k)}_LR(T_c-\epsilon_k).
\end{gather}

Using (\ref{eq:phi1}) and (\ref{eq:phi2}),
$\textrm{E}\{(a_m^{(k)})^2|\epsilon_k\}$ can be expressed from
(\ref{eq:mai_asyc_ap2}) as
\begin{align}\label{eq:son_bu}\nonumber
\textrm{E}\{(a_m^{(k)})^2|\epsilon_k\}&=\frac{1}{N_c}\sum_{j=0}^{L-1}\left(\sum_{i=1}^{j}\beta_i[\alpha^{(k)}_{i+L-j-1}R(T_c-\epsilon_k)+\alpha^{(k)}_{i+L-j}R(\epsilon_k)]+\beta_{j+1}\alpha^{(k)}_{L}R(T_c-\epsilon_k)\right)^2\\
&+\frac{1}{N_c}\sum_{j=0}^{L-1}\left(\sum_{i=1}^{j}\alpha^{(k)}_i[\beta_{i+L-j-1}R(\epsilon_k)+\beta_{i+L-j}R(T_c-\epsilon_k)]+\alpha_{j+1}^{(k)}\beta_LR(\epsilon_k)\right)^2.
\end{align}
Also, due to the polarity codes, the correlation terms are zero.
That is, $\textrm{E}\{a_m^{(k)}a_n^{(k)}\}=0$ for $m\ne n$. Then,
from the central limit argument in \cite{meas}, we see that
$a^{(k)}$ in (\ref{eq_MAI_MP}), conditioned on $\epsilon_k$,
converge to the distribution given in Lemma
\ref{lem:asyc1_MAI_MP}.

\subsection{Proof of Lemma \ref{lem:asyc2_MAI_MP}}\label{app:asyc2_MAI_MP}

Consider $(N_u-1)$ interfering users, each with bit energy $E$.
Then, the total MAI $a=\sum_{k=2}^{N_u}a^{(k)}$ is the sum of
$(N_u-1)$ i.i.d. random variables, where
$a^{(k)}=\sqrt{\frac{E}{N_f}}\sum_{m=iN_f}^{(i+1)N_f-1}a_m^{(k)}$.
Using the results in Appendix \ref{app:asyc1_MAI_MP}, namely,
$\textrm{E}\{a_m^{(k)}\}=0$, $\textrm{E}\{a_m^{(k)}a_n^{(k)}\}=0$
for $m\ne n$ and (\ref{eq:son_bu}), we obtain
\begin{gather}
\frac{1}{\sqrt{N_u-1}}\sum_{k=2}^{N_u}a^{(k)}\sim\mathcal{N}\left(0\,,\,\frac{E}{N_c}\textrm{E}\{\sigma^2_{MAI,k}(\epsilon_k)\}\right),
\end{gather}
as $N_u\longrightarrow\infty$, where
$\textrm{E}\{\sigma^2_{MAI,k}(\epsilon_k)\}$ can be obtained as in
(\ref{eq:exp_value}) from (\ref{eq:son_bu}) using the fact that
$\epsilon_k\sim\mathcal{U}[0,T_c)$.

\end{document}